\documentclass[preprint,lettersize,journal]{IEEEtran}
\usepackage{amsmath,amsfonts}
\usepackage{algorithmic}
\usepackage{array}
\usepackage[caption=false,font=normalsize,labelfont=sf,textfont=sf]{subfig}
\usepackage{textcomp}
\usepackage{fancyhdr} 
\usepackage{stfloats}
\usepackage{orcidlink}
\usepackage{url}
\usepackage{verbatim}
\usepackage{graphicx}
\usepackage{hyperref}
\def\BibTeX{{\rm B\kern-.05em{\sc i\kern-.025em b}\kern-.08em
    T\kern-.1667em\lower.7ex\hbox{E}\kern-.125emX}}
\usepackage{balance}

\begin{document}

\title{Statistical Study of Sensor Data and Investigation of ML-based Calibration Algorithms for Inexpensive Sensor Modules: Experiments from Cape Point}
\author{Travis Barrett\orcidlink{0000-0001-9040-8621}, and Amit Kumar Mishra\orcidlink{0000-0001-6631-1539}, \textit{Senior Member}, \textit{IEEE}
\thanks{This work has been supported with funding from Sentech Soc Ltd, South Africa.}}
\maketitle

\begin{abstract}
In this paper we present the statistical analysis of data from inexpensive sensors. We also present the performance of machine learning algorithms when used for automatic calibration such sensors. In this we have used low-cost Non-Dispersive Infrared CO$_2$ sensor placed at a co-located site at Cape Point, South Africa (maintained by Weather South Africa). 
The collected low-cost sensor data and site truth data are investigated and compared. We compare and investigate the performance of Random Forest Regression, Support Vector Regression, 1D Convolutional Neural Network and 1D-CNN Long Short-Term Memory Network models as a method for automatic calibration and the statistical properties of these model predictions. 
In addition, we also investigate the drift in performance of these algorithms with time. 
\end{abstract}


\section{Introduction}

This paper is an extension of our work \cite{conference2023} on the use of cost-effective agile sensor platforms and the improvement of the collected data using various machine learning implementations and makes reuse of the work presented.

In the age of the Internet of Things (IoT), many projects are being created \cite{lee2015internet} to meet the ever-expanding web of systems that are all interconnected. 
With the goal of many of these projects to be as cost-effective as possible, low-cost sensors are often used. These sensors have varying levels of accuracy and performance characteristics that are influenced by the environment. Due to this, the performance of these sensors can vary heavily on their location which can produce erroneous data. Without a thorough investigation and methodology around sensor calibration, usually the data collected from the sensors is not reliable. For example, Bittner et al  \cite{bittner2022performance} discussed how they got their sensors calibrated to a high standard and then those were installed in Malawi. However, they observed how quickly the quality of the data decreased from the sensor network and eventually became mostly unusable.  This is a major pain-point in the current generation of ubiquitous sensing. It has been well described by Motlagh et al~\cite{Motlagh2020MassiveScaleAirQuality} that a high spatial density is required to effectively monitor the air quality of an area. Having a cost-effective solution that can be tailored to each location would allow for wide-scale monitoring to be more accessible. Motlagh et al~\cite{Motlagh2020MassiveScaleAirQuality} also state that large scale air quality monitoring would require opportunistic calibration transfer as the scale of deployments renders manual calibration infeasible. 

Calibration can be a complicated and challenging process. 
Laboratory-based calibration is a consistent method to properly calibrate sensors. 
However, these laboratory conditions are not fully representative of the wide array of conditions the sensors may face \cite{rai2017end}. 
This will cause deviations between the calibrated and the observed data. 
To re-calibrate the sensors, as they will inevitably drift over time, the sensor will need to be returned to the laboratory to be calibrated again. The alternative to this is to take reference calibration equipment to the sensors in the field. \textcolor{black}{These calibration processes are one-point, two-point or span calibration and multi-point curve fitting calibration.} This is a difficult and expensive method to calibrate the sensors in place as it will require a dedicated team to calibrate a sensor network. This process will not scale up to a large and/or remote sensor networks such as the ones seen in the work done by Hepworth and Mishra \cite{hepworth2023analysis}. In order to calibrate these sensors a reference sensor is required. These sensors are typically very expensive where accuracy of measurement is paramount. For instance, the Picarro G2401 sensor system used by the South African Weather Service in their Cape Point climate measurement station is costly and is not easily moved around. 
While it may not be possible to remove the need for reference equipment entirely, a longer time between the manual calibrations of sensors will greatly reduce the cost and staff requirements for a given size of sensor network. 
The longer the sensors can have their data accurately calibrated through software methods, the lower the maintenance requirements of the network.  

As shown in Postolache~\cite{postolache2009smart} the ability for low-cost sensors networks to be created has been around for many years. 
Making a system today gives the ability to use improved sensor technologies and micro-controllers to create more flexible and accurate nodes. With the greater availability of communication technologies combined with the work done to improve calibration of these sensor nodes, this paper shows the potential of inexpensive and agile platform (one of which was the focus of our conference paper \cite{conference2023}) for wide-scale deployment.

Machine learning (ML) is an attractive prospect for calibration and management of data produced by sensor systems. \textcolor{black}{An example of machine learning, specifically support vector regression, being used to improve the performance of a sensor can be seen in the work done by Cheng et al \cite{cheng2016prediction} where machine learning was suggested to improve the accuracy of a differential absorption non-dispersive infrared sensor. Another example is the work doe by Zimmerman et al \cite{zimmerman2018machine} where a Random Forest model was used to calibrate and improve air quality sensor performance in an air quality monitoring platform.} Machine learning can be used as a tool to group or even predict data based on models trained on the collected data. While this is useful and appears to be functional on the surface, in the case of data prediction, how can we verify that these models are actually performing as expected and for how long are these models viable? 
Secondly, even though there is a growing body of work on the use of ML for sensor-calibration, hardly any look at the reliability of these algorithms over time. I.e. what is the drift of the performances of these algorithms over time? 
Lastly, understanding the data is the key to the performance of any data processing algorithm. Statistical analysis of data from inexpensive sensors is something not dealt with extensively in open literature. 

In our paper, we endeavour to bridge all the above gaps. 
The work has four main novelties. Firstly, we present low-cost non-dispersive infrared (NDIR) \textcolor{black}{CO$_2$} sensor data set from a co-located site with ground truth data coming from a metrological standard high-fidelity (Hi-Fi) sensor system maintained by Weather South Africa at their Cape Point weather station. It can be noted that the choice of sensing \textcolor{black}{CO$_2$} was based on the available sensors at the Weather SA station. 
The work is generic and can be expanded to any other sensor. 
This data contains the drift of the sensor as well as different failure states (of our inexpensive sensor module). 
Secondly, we shall present an analysis of the performance of different machine learning models based on the size of training data and the time between training and predication. 
Thirdly, we shall present statistical and performance metrics of these ML models to determine their performance over time. 
Lastly, we shall present a thorough statistical analysis of the data (both from our inexpensive sensor module as well as from the Hi-Fi sensors from the weather station). In this, we shall study the Gaussianity as well as the ergodicity of the data. 
 

To determine the performance of the different machine learning models used, a set of metrics was chosen to compare the characteristics of the predictions produced and the truth data. These metrics, listed in Table \ref{tab:metrics}, allow for the comparison between the data considering absolute error as well as fit and statistical similarity. 
The aim of having these metrics is to determine how effective the models are at predicting the truth data from the collected sensor data and in what way the predicted data best fits the truth data.

\begin{table}[!h]
\caption{Table of metrics used for comparison of machine learning model performance.}
\normalsize
\centering
\begin{tabular}{|l|}
\hline
\textbf{Metrics}                          \\ \hline
Mean Absolute Error             \\
Accuracy (\%)                   \\
R$^2$                           \\
Pearson Correlation Coefficient \\
Kullback-Liebler Divergence     \\
Jenson-Shannon Divergence       \\ \hline
\end{tabular}%

\label{tab:metrics}
\end{table}


The rest of the paper is organized as follows. 
Section II describes the experimental setup. The following section presents the statistical analysis of the collected data. 
Section IV presents the machine learning models for calibration of the sensor data. 
The following section presents the statistical and ergodicity study of the performance of the machine learning models. Section VI concludes the paper. 
 
\section{Low Cost NDIR Sensor Data}
In this section, we shall describe the design of the low-cost sensor module to measure \textcolor{black}{CO$_2$}. As we have mentioned before, the choice of \textcolor{black}{CO$_2$} was based on the availability of Hi-Fi sensors in the Weather SA station. 
The sensor platform developed in our work is flexible and can easily be modified to host other kinds of low-cost sensors. 

\subsection{Experimental Setup}
\subsubsection{System Design and Information}
In order to collect data, a measurement system was created. The platform is based on Espressif's ESP32 System on a Chip (SoC) \cite{esp32_Datasheet} due to its ease of access and cost-effective features. This allowed for creating a network connected, agile, measurement platform with the ability to connect to multiple sensors. The platform, being designed in South Africa, was designed with an on-board Uninterruptible Power Supply (UPS) so that the platform could continue to monitor in the event the mains power was disabled, as is the case during load-shedding. The platform is currently designed to break out to two hardware Universal Asynchronous Receiver/ Transmitter (UART) headers for two separate sensors. These two sensors were an MH-Z19C CO$_2$ Non-dispersive Infrared (NDIR), measuring CO$_2$ parts per million (ppm), and a ZH03B laser-scattering Particulate Matter (PM) sensor, measuring micrograms per cubic metre \textcolor{black}{($\mu$g/m$^3$)}, both made by Winsen Electronics Technology Co. Both of these sensors only report integer values and do not have a very high level of accuracy, \textcolor{black}{±50 ppm + 5\%} reading value and ±15 \textcolor{black}{$\mu$g/m$^3$} respectively, when compared to that of the reference equipment. The platform is not limited to these sensors as the board can support any hardware compatible sensor with the required software changes. This platform has been being operated since 08/12/2021 at the University of Cape Town and was deployed at the South African Weather Service Cape Point measurement station since 12/09/2022.
\begin{figure}[!h]
    \centering
    \includegraphics[width=5cm]{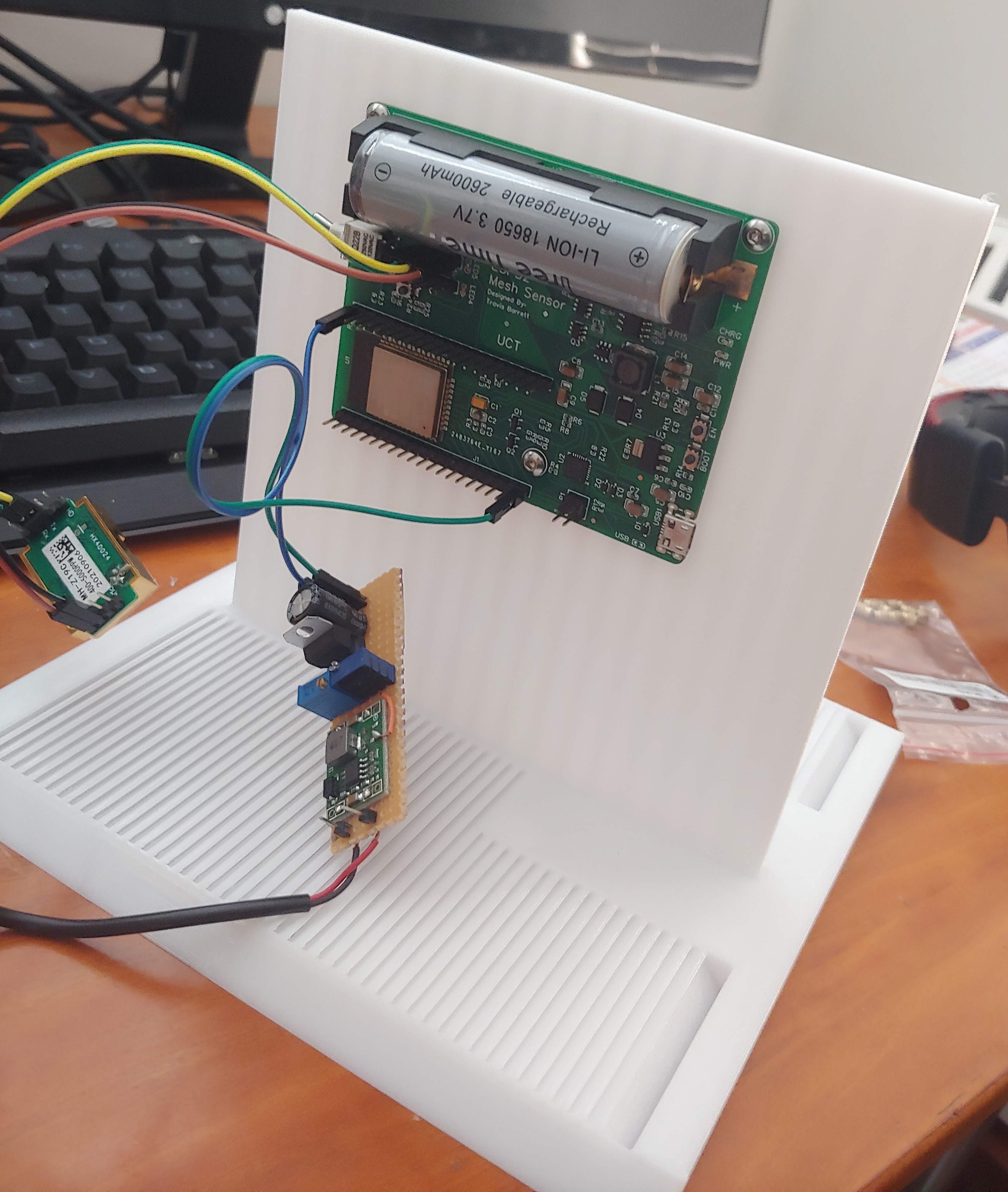}
    \caption{This image shows the sensor system connected to the backbone of its purpose built, double-louvered Stevenson screen, enclosure during preparation for deployment at Cape Point, South Africa.}
    \label{fig:sensor}
\end{figure}

\subsubsection{Co-Location Site and Equipment}
The South African Weather Service Cape Point measurement station is located in the Cape Point nature reserve in Cape Town, South Africa. This measurement site works with the Global Monitoring Laboratory (GML) of the National Oceanic and Atmospheric Administration who specialize in research into greenhouse gas and carbon cycle feedbacks, changes in clouds, aerosols, and surface radiation, and recovery of stratospheric ozone \cite{NOAA}. In this measurement site, the Picarro G2401 is used as the ground truth for the calibration of the test system and has an accuracy of 50, 20, or 10 parts per billion (ppb) depending on the chosen sample rate \cite{picarroG2401}. It is important to note that this measurement station is measuring atmospheric concentrations and is doing so by taking air from the top of a mast for their measurements. Our measurement platform was placed in an adjoining open-air room at the measurement facility and is subsequently more susceptible to local condition changes due to its location.

\subsubsection{\textcolor{black}{Error sources of NDIR CO$_2$ Sensors}}
\textcolor{black}{As seen in the work done by Marinov et al \cite{Marinov2015NDIRSensorAccuracy}, there are environmental factors that can cause errors in NDIR CO$_2$ sensors. One such factor is that the sensor is detecting the molecular density of CO$_2$ in the measurement chamber. As the molecular density of a gas changes with pressure, this can cause a measurement error if not properly compensated for. The temperature of the gas is another source of error as it forms part of the ideal gas equation. This is shown to produce relative errors of $\pm$35\% in the testing done by Marinov et al \cite{Marinov2015NDIRSensorAccuracy}. Humidity is another source of error as the water molecules in the measurement gas can be cross-detected by the NDIR sensor. However this is less severe than that of temperature and pressure variations. These random environmental factors, if not considered when taking measurements with NDIR sensors can greatly impact the accuracy of the readings. The infrared emitter in the NDIR sensor itself is a source of systematic error as over time the emitter degrades and produces a reduced output. This, if not compensated for, will cause the sensor readings to drift as the baseline reference output is changing. Additionally the detection element may degrade over time causing drift. Another source of error is found as the systematic quantisation error for systems with digital outputs. However, these are relatively negligible when compared to the other sources of errors.}

\subsection{Results}
Due to the deployment of this system, a data set has been created spanning 12/09/2022 to 31/10/2023 at the time of writing. The set contains interruptions due to the nature of the local power supply issues that we experience in South Africa. 
However, the data collected is able to be compared to a highly accurate reference onsite as the sensors are co-located. 

\subsection{Discussion}
The creation of this data set forms the basis for the rest of the work in this paper. Our goal to improve the performance of these low-cost sensors was a result of creating and operating this platform. The data collected is not uninterrupted which is a reality of these low-cost sensor platforms as well as the environment in which we are testing the platform. South Africa has experiencing scheduled rolling blackouts and as a result critical infrastructure may not function as expected. While attempts were made to mitigate these effects, they were not entirely avoided. 
It can be noted that the situation in many parts of the Global South is similar. Hence, our study has a wider pertinence. 
In future we wish to continue co-location at this site while adding other sites to allow for collection from different sensors as well as more diverse sites to further increase the size and scope of our data.

\section{Analysis of Collected Data}
In this section the statistical properties of the data will be investigated.

\subsection{Experimental Setup}
The data has been stored in files containing readings taken every ten seconds by the test sensor system. Due to a variety of reasons, there are missing values in the data set which need to be accounted for when doing analysis on, and using, this collected data. The data used in this section are values that have been matched within a 60-second window of their corresponding values in the data from Cape Point's one minute interval data. This will allow for comparing and training on data that is time-matched. \textcolor{black}{This is achieved by locating values from the collected sensor data within the 60-second window around the truth data point. If six values are found they are paired with the corresponding truth data point. If there are more than three values but less than six, the remaining values are populated with the mean of the existing values present in the time window and paired with the corresponding truth data point. If three or less values are present they are discarded as well as the truth data point. In the case of more than six values, the first six are paired with the truth data point.} \textcolor{black}{This has resulted in the creation of three data sets with corresponding matched truth data.} \textcolor{black}{Data Set} 1 spans from 09/09/2022 to 22/10/2022 and \textcolor{black}{Data Set 2} spans from 22/01/2023 to 02/02/2023. \textcolor{black}{Data Set} 3 will not be used for analysis in this section. The spans of the data sets can be seen in Table \ref{tab:Data-Sets} These sets were chosen as they had the most consistent periods of matched values.
\begin{table}[!h]
\caption{\textcolor{black}{Table listing the time-span of collected data sets}}
\label{tab:Data-Sets}
\centering
\begin{tabular}{|lll|}
\hline
\textbf{Data Set}                & \textbf{Start Date} & \textbf{End Date}   \\ \hline
\multicolumn{1}{|l|}{1} & 09/09/2022 & 22/10/2022 \\
\multicolumn{1}{|l|}{2} & 22/01/2023 & 02/02/2023 \\
\multicolumn{1}{|l|}{3} & 18/03/2023 & 25/10/2023 \\ \hline
\end{tabular}
\end{table}

One of the important statistical points of interest is whether the data is Gaussian or not. To test this, for each set, the data collected from the test system and the Cape Point measurement site are split into two equal parts and plotted as a histogram. 
This process is also done with four equal divisions of the data for the same reason. For each collected data slice, we apply a Shapiro-Wilk test and Lilliefors' test to determine the Gaussianity of the data. 
We will also calculate the covariance matrices of \textcolor{black}{Data Set 1 and 2 and truth data set pair as well as Data Set 1 with Data Set 2.}

\subsection{Results}
\begin{figure}[!h]
    \centering
    \includegraphics[width=\columnwidth]{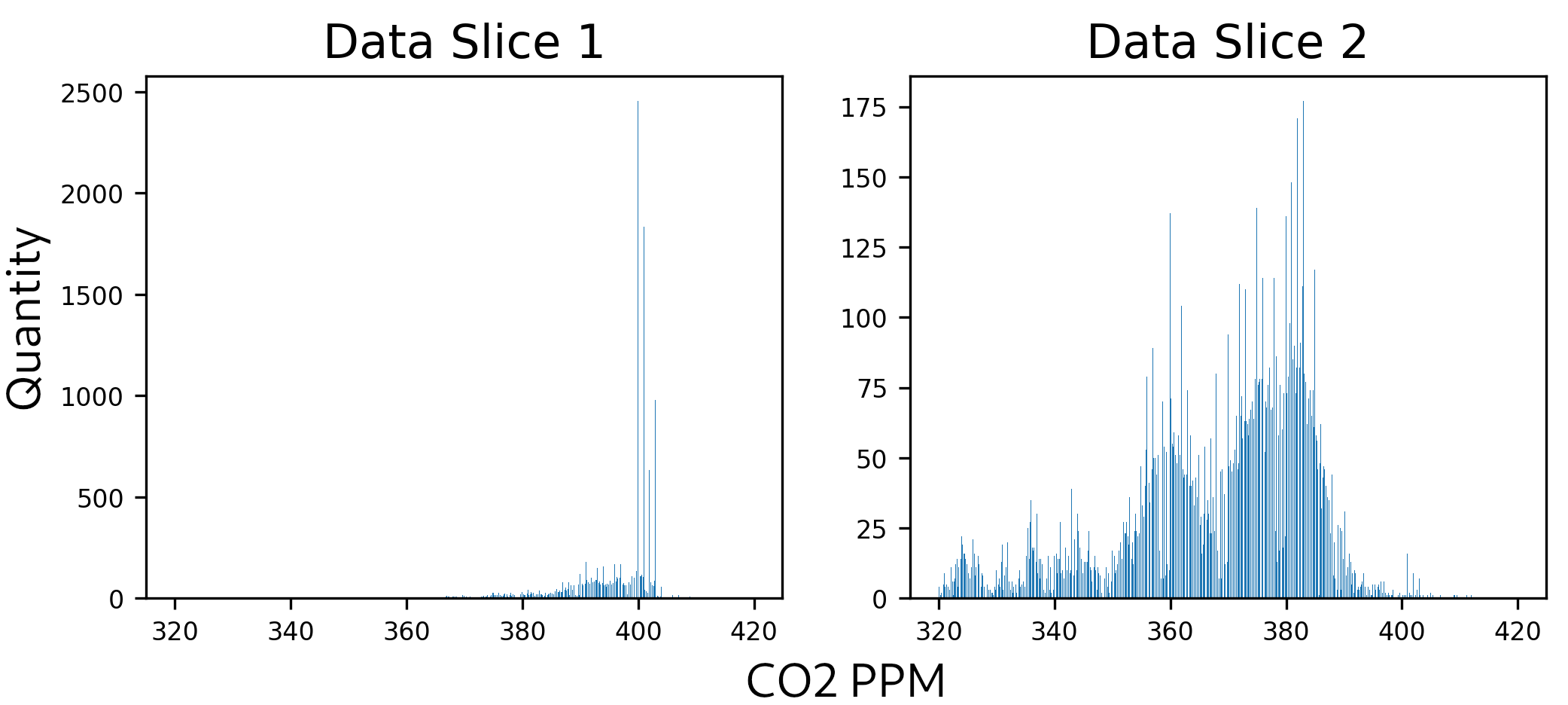}
    \caption{Histogram of \textcolor{black}{Data Set} 1 split into two slices for comparison.}
    \label{fig:Hist-2-set1-features}
\end{figure}

\begin{figure}[!h]
    \centering
    \includegraphics[width=\columnwidth]{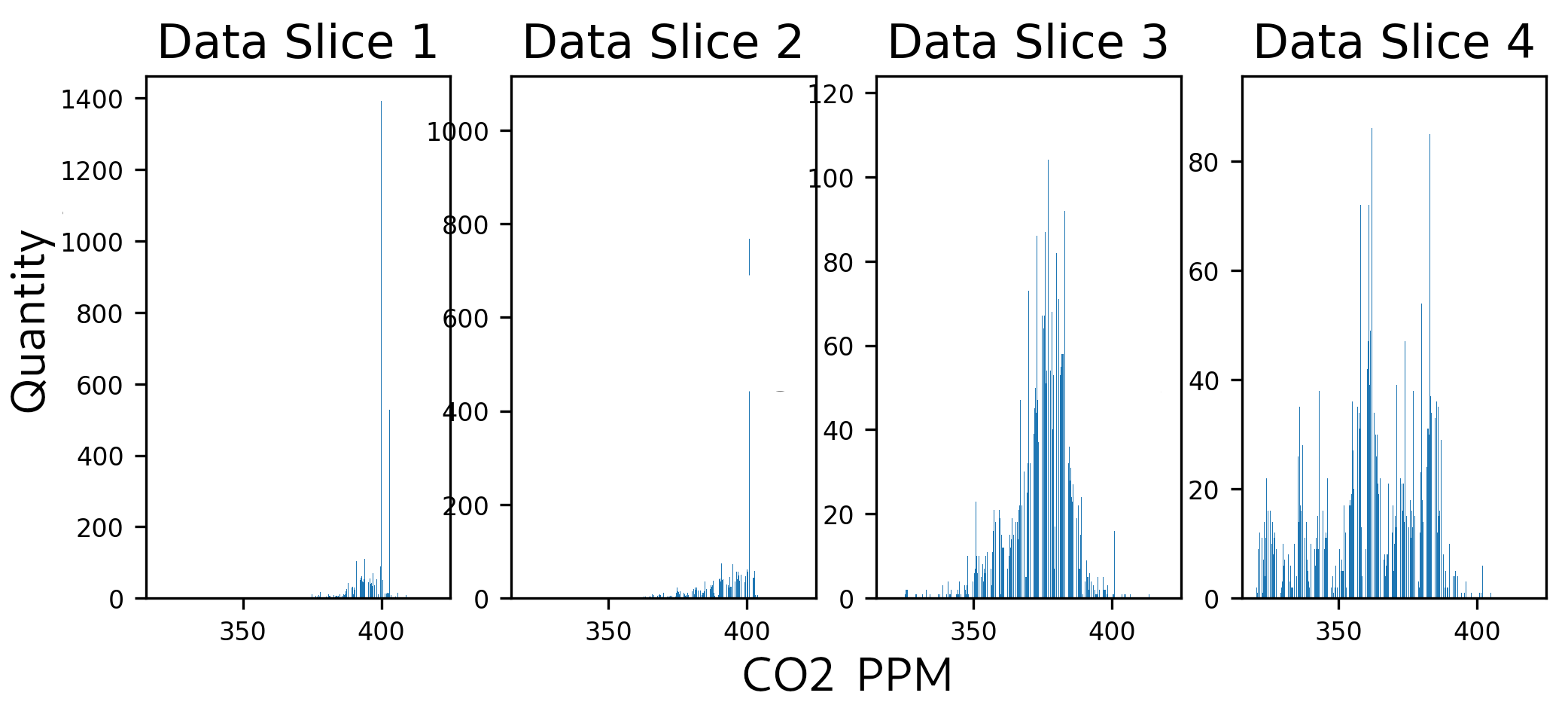}
    \caption{Histogram of \textcolor{black}{Data Set} 1 split into four slices for comparison.}
    \label{fig:Hist-4-set1-features}
\end{figure}

\begin{figure}[!h]
    \centering
    \includegraphics[width=\columnwidth]{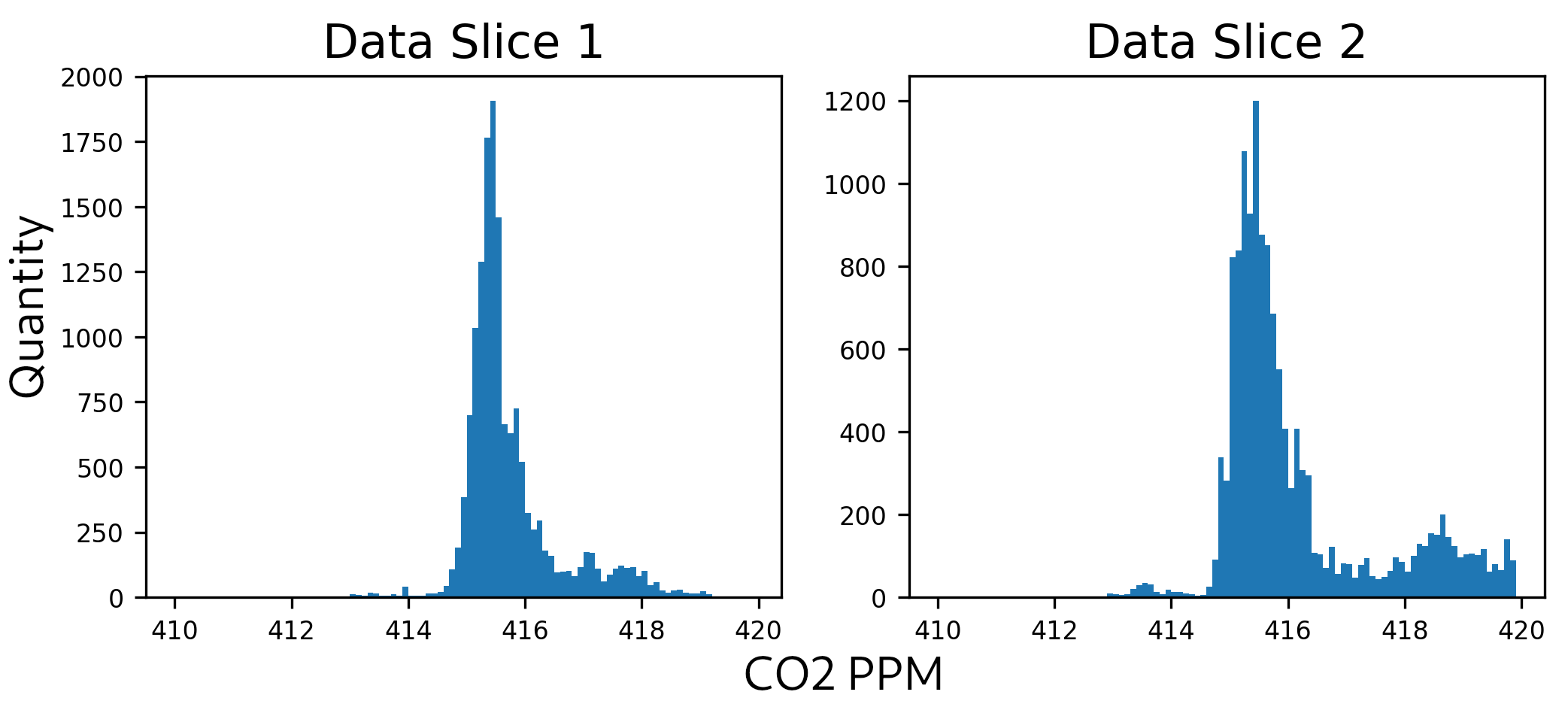}
    \caption{Histogram of \textcolor{black}{Data Set} 1 truth data split into two slices for comparison.}
    \label{fig:Hist-2-set1-labels}
\end{figure}

\begin{figure}[!h]
    \centering
    \includegraphics[width=\columnwidth]{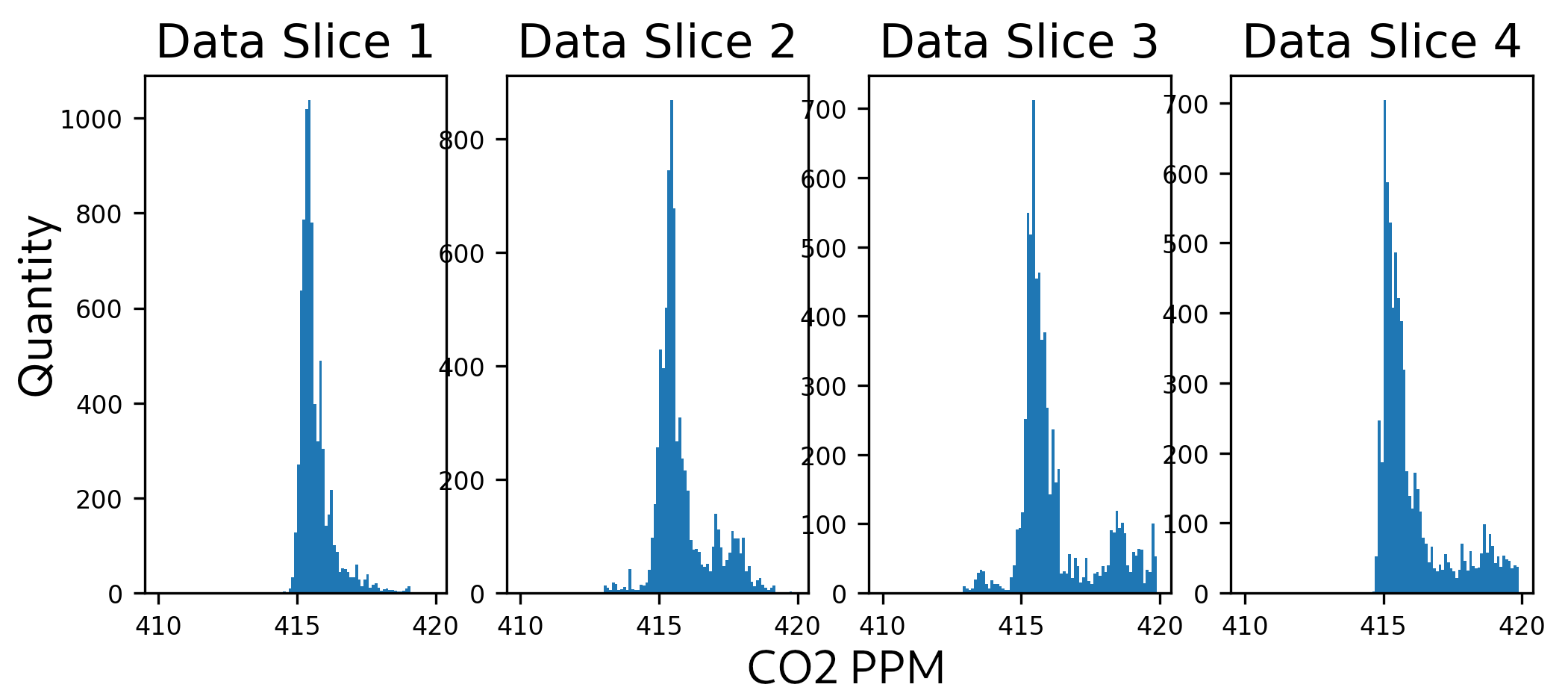}
    \caption{Histogram of \textcolor{black}{Data Set} 1 truth data split into four slices for comparison.}
    \label{fig:Hist-4-set1-labels}
\end{figure}

\begin{figure}[!h]
    \centering
    \includegraphics[width=\columnwidth]{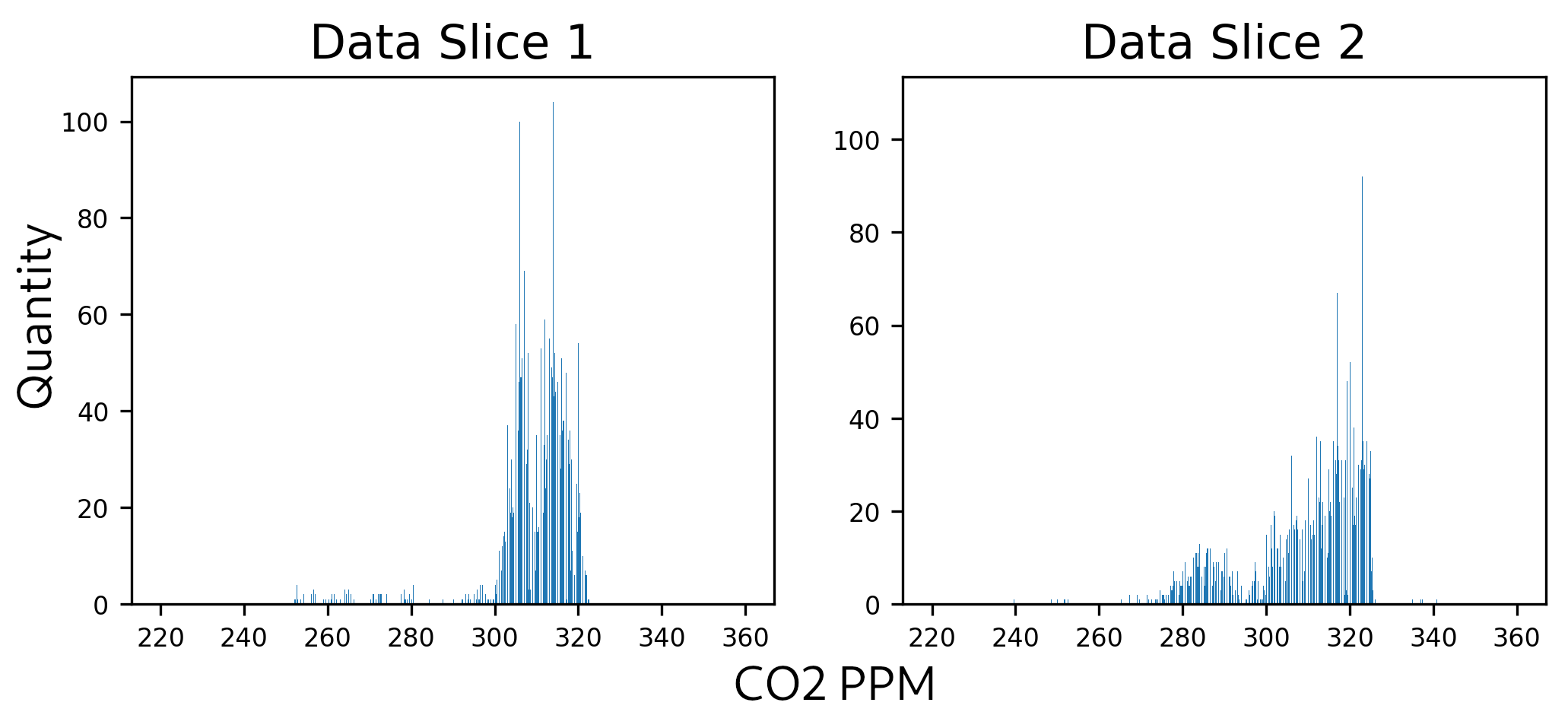}
    \caption{Histogram of \textcolor{black}{Data Set} 2 split into two slices for comparison.}
    \label{fig:Hist-2-set2-features}
\end{figure}

\begin{figure}[!h]
    \centering
    \includegraphics[width=\columnwidth]{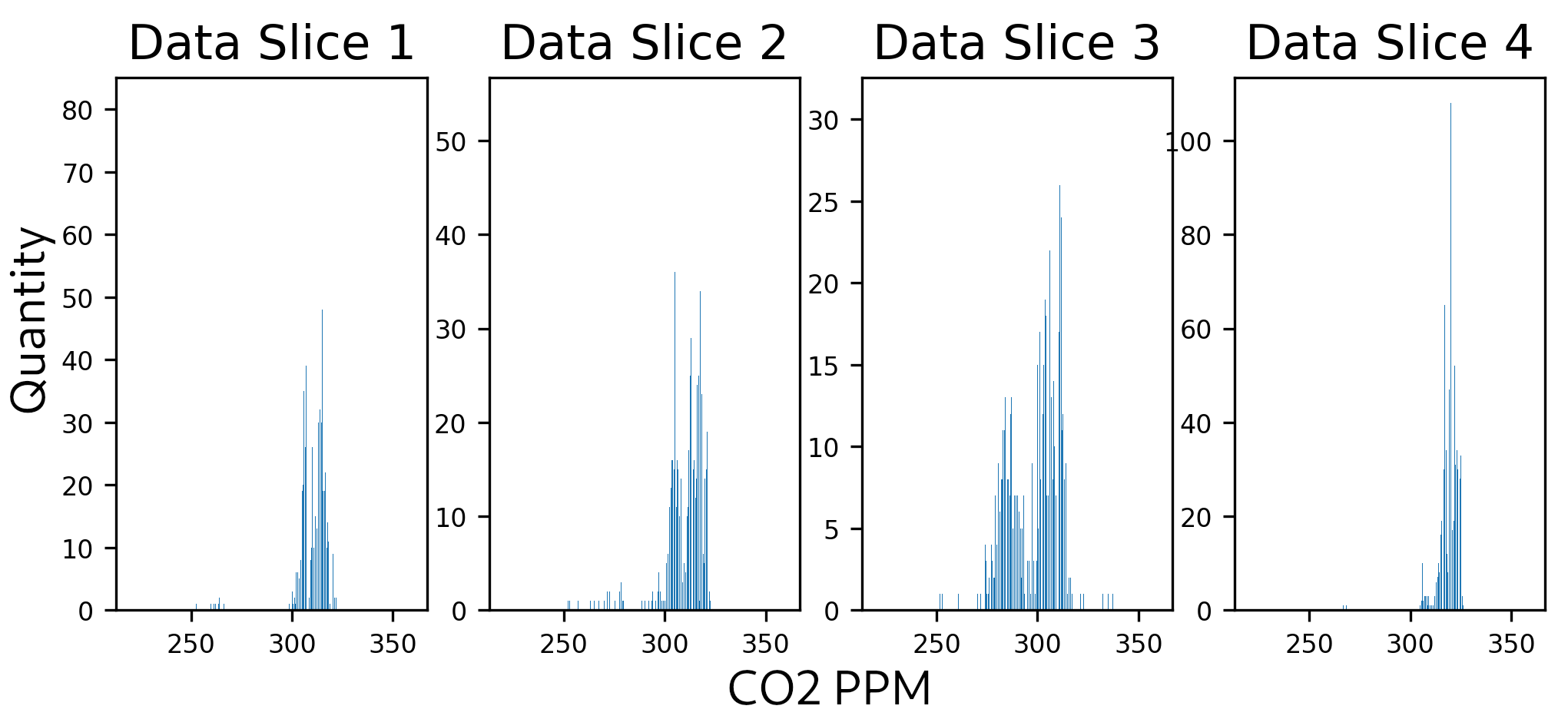}
    \caption{Histogram of \textcolor{black}{Data Set} 2 split into four slices for comparison.}
    \label{fig:Hist-4-set2-features}
\end{figure}

\begin{figure}[!h]
    \centering
    \includegraphics[width=\columnwidth]{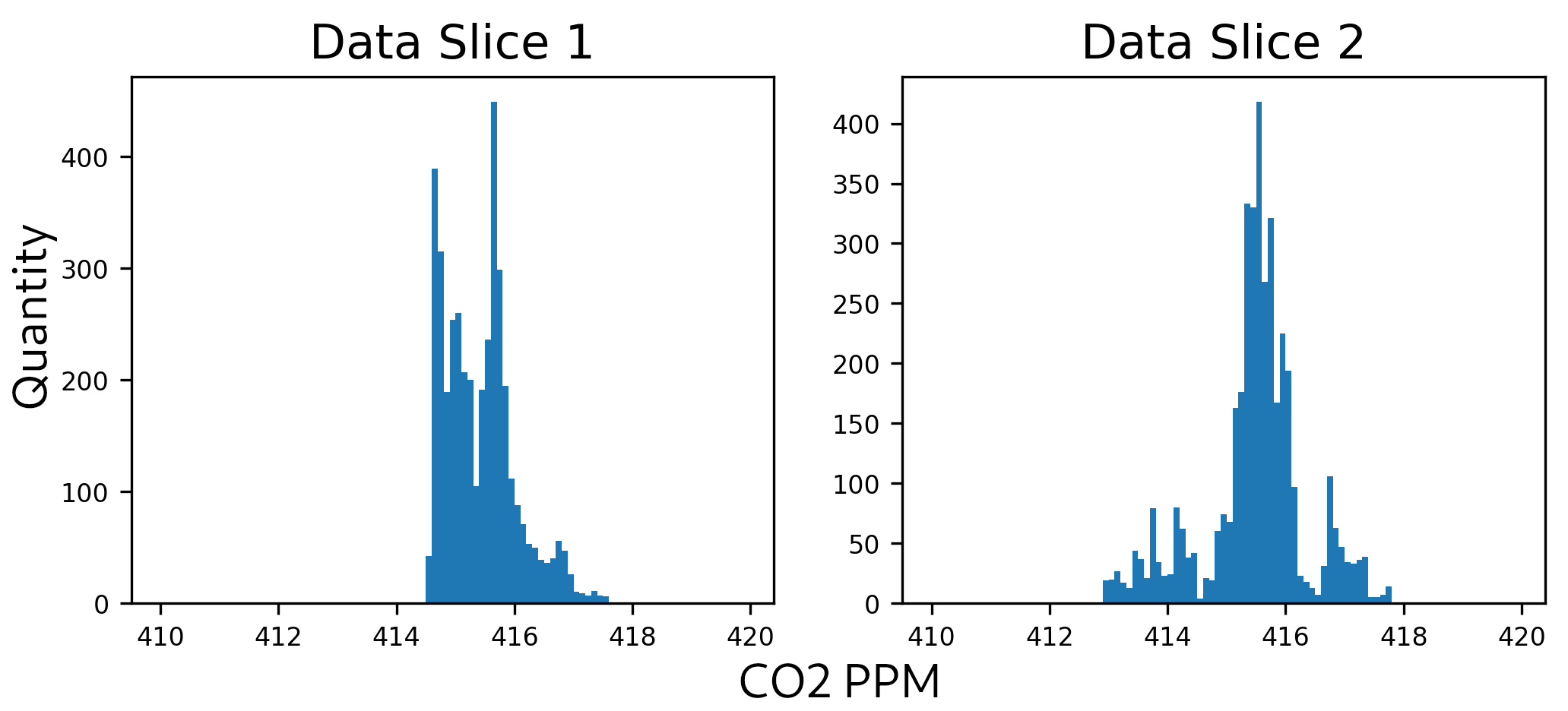}
    \caption{Histogram of \textcolor{black}{Data Set} 2 truth data split into two slices for comparison.}
    \label{fig:Hist-2-set2-labels}
\end{figure}

\begin{figure}[!h]
    \centering
    \includegraphics[width=\columnwidth]{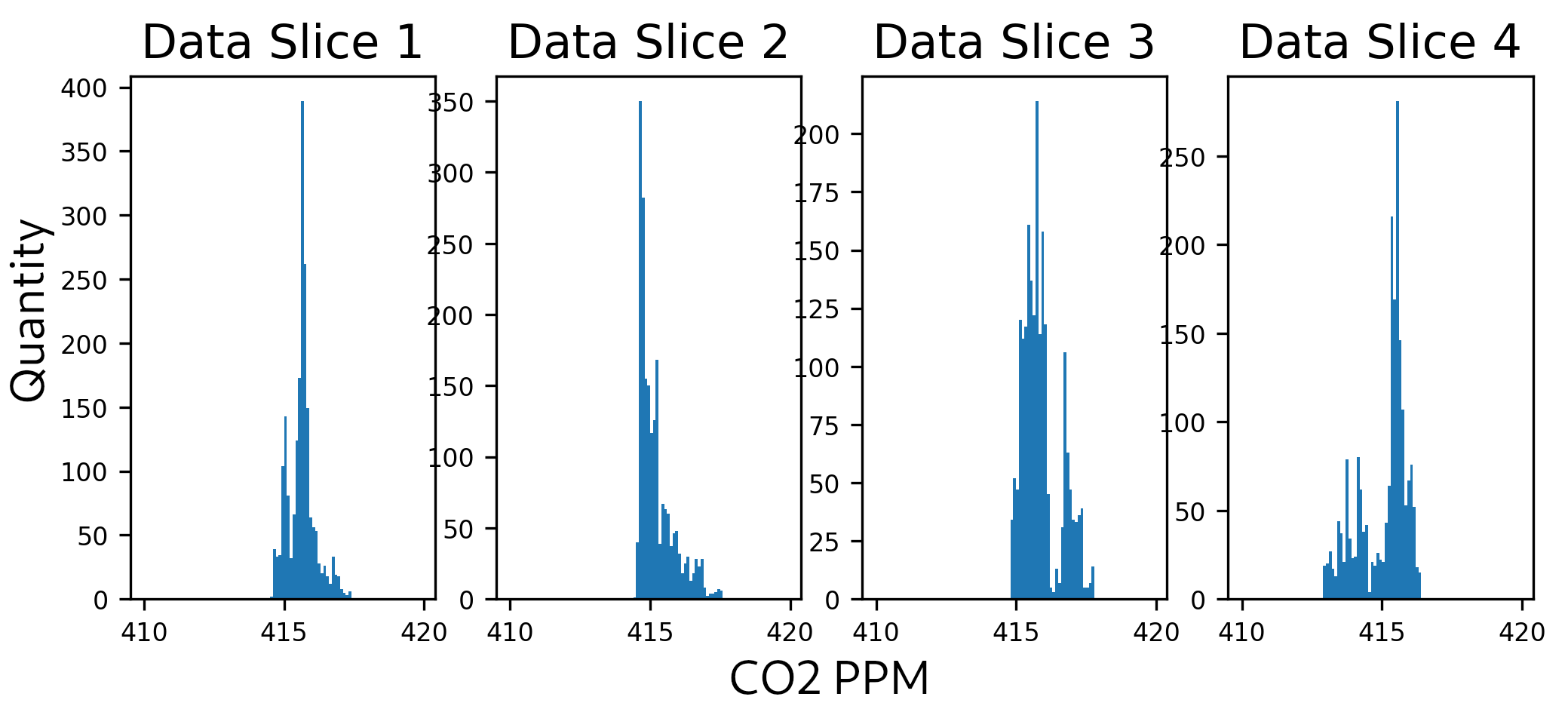}
    \caption{Histogram of \textcolor{black}{Data Set} 2 truth data split into four slices for comparison.}
    \label{fig:Hist-4-set2-labels}
\end{figure}

The results from the experiments are presented, \textcolor{black}{ in Figures~\ref{fig:Hist-2-set1-features}, \ref{fig:Hist-4-set1-features}, \ref{fig:Hist-2-set1-labels}, \ref{fig:Hist-4-set1-labels}, \ref{fig:Hist-2-set2-features}, \ref{fig:Hist-4-set2-features}, \ref{fig:Hist-2-set2-labels} and \ref{fig:Hist-4-set2-labels},} showing the distributions of the data when viewed in slices. 
The Shapiro-Wilk and Lilliefors' test can be seen in Tables \ref{tab:Shap-wilk-2feat-2}-\ref{tab:Lilliefors-2label-4}. The following matrices are the covariance matrices of the individual collected data sets and their matched truth data as well as the covariance matrix of the two collected data sets.\\
\textbf{Covariance Matrix of \textcolor{black}{Data Set 1 and Matched Truth Data Set:}}
\[
\begin{bmatrix}
    418.56579704&-1.48743692 \\
    -1.48743692&2.8509566 
\end{bmatrix}
\]
\textbf{Covariance Matrix of \textcolor{black}{Data Set 2 and Matched Truth Data Set:}}
\[
\begin{bmatrix}
    174.43116894&-0.77310627 \\
    -0.77310627&0.63195665
\end{bmatrix}
\]
\textbf{Covariance Matrix of \textcolor{black}{Data Set 1 and Data Set 2:}}
\[
\begin{bmatrix}
    50.97994586&11.70838145 \\
    11.70838145&174.43116894
\end{bmatrix}
\]


\begin{table}[!h]
\caption{Shapiro-Wilk test performed on \textcolor{black}{Data Set} 2. The data has been split into two slices that have been tested for their normality.}
\normalsize
\centering
\begin{tabular}{|l|l|l|}
\cline{1-3}
\textbf{Slice Number} & \textbf{Statistic}          & \textbf{P-Value}  \\ \cline{1-3}
1            & 0.585 & 0.000   \\ 
2            & 0.878 & 0.000                      \\ \cline{1-3}
\end{tabular}%
\label{tab:Shap-wilk-2feat-2}
\end{table}

\begin{table}[!h]
\caption{Shapiro-Wilk test performed on \textcolor{black}{Data Set} 2. The data has been split into four slices that have been tested for their normality.}
\normalsize
\centering
\begin{tabular}{|l|l|l|}
\hline
\textbf{Slice Number} & \textbf{Statistic}          & \textbf{P-Value}                 \\ \hline
1            & 0.475 & 0.000   \\ 
2            & 0.682 & 0.000   \\ 
3            & 0.898 & 0.000   \\ 
4            & 0.668 & 0.000                      \\ \hline
\end{tabular}%
\label{tab:Shap-wilk-2feat-4}
\end{table}

\begin{table}[!h]
\caption{Lilliefors' test performed on \textcolor{black}{Data Set} 2. The data has been split into two slices that have been tested for their normality.}
\normalsize
\centering
\begin{tabular}{|l|l|l|}
\cline{1-3}
\textbf{Slice Number} & \textbf{Statistic}          & \textbf{P-Value}  \\ \cline{1-3}
1            & 0.190 & 0.001   \\ 
2            & 0.126 & 0.001                     \\ \cline{1-3}
\end{tabular}%
\label{tab:Lilliefors-2feat-2}
\end{table}

\begin{table}[!h]
\caption{Lilliefors' test performed on \textcolor{black}{Data Set} 2. The data has been split into four slices that have been tested for their normality.}
\normalsize
\centering
\begin{tabular}{|l|l|l|}
\cline{1-3}
\textbf{Slice Number} & \textbf{Statistic}          & \textbf{P-Value}                 \\ \cline{1-3}
1            & 0.247 & 0.001   \\ 
2            & 0.152 & 0.001   \\ 
3            & 0.127 & 0.001   \\ 
4            & 0.138 & 0.001   \\ \cline{1-3}
\end{tabular}%
\label{tab:Lilliefors-2feat-4}
\end{table}

\begin{table}[!h]
\caption{Shapiro-Wilk test performed on \textcolor{black}{Data Set} 2 matched truth data. The data has been split into two slices that have been tested for their normality.}
\normalsize
\centering
\begin{tabular}{|l|l|l|}
\cline{1-3}
\textbf{Slice Number} & \textbf{Statistic}          & \textbf{P-Value}  \\ \cline{1-3}
1            & 0.943 & 0.000   \\ 
2            & 0.834 & 0.000                      \\ \cline{1-3}
\end{tabular}%
\label{tab:Shap-wilk-2label-2}
\end{table}

\begin{table}[!h]
\caption{Shapiro-Wilk test performed on \textcolor{black}{Data Set} 2 matched truth data. The data has been split into four slices that have been tested for their normality.}
\normalsize
\centering
\begin{tabular}{|l|l|l|}
\hline
\textbf{Slice Number} & \textbf{Statistic}          & \textbf{P-Value}                 \\ \hline
1            & 0.950 & 0.000   \\ 
2            & 0.844 & 0.000   \\ 
3            & 0.630 & 0.000   \\ 
4            & 0.885 & 0.000                      \\ \hline
\end{tabular}%
\label{tab:Shap-wilk-2label-4}
\end{table}

\begin{table}[!h]
\caption{Lilliefors' test performed on \textcolor{black}{Data Set} 2 matched truth data. The data has been split into two slices that have been tested for their normality.}
\normalsize
\centering
\begin{tabular}{|l|l|l|}
\cline{1-3}
\textbf{Slice Number} & \textbf{Statistic}          & \textbf{P-Value}  \\ \cline{1-3}
1            & 0.075 & 0.001   \\ 
2            & 0.151 & 0.001                     \\ \cline{1-3}
\end{tabular}%
\label{tab:Lilliefors-2label-2}
\end{table}

\begin{table}[!h]
\caption{Lilliefors' test performed on \textcolor{black}{Data Set} 2 matched truth data. The data has been split into four slices that have been tested for their normality.}
\normalsize
\centering
\begin{tabular}{|l|l|l|}
\cline{1-3}
\textbf{Slice Number} & \textbf{Statistic}          & \textbf{P-Value}                 \\ \cline{1-3}
1            & 0.110 & 0.001   \\ 
2            & 0.160 & 0.001   \\ 
3            & 0.153 & 0.001   \\ 
4            & 0.221 & 0.001   \\ \cline{1-3}
\end{tabular}%
\label{tab:Lilliefors-2label-4}
\end{table}

\subsection{Discussion}
It is clear from the comparison of the data collected by the sensor system, seen in Figures \ref{fig:Hist-2-set1-features} and \ref{fig:Hist-2-set2-features}, to that of the Cape Point truth data, seen in Figures \ref{fig:Hist-2-set1-labels} and \ref{fig:Hist-2-set2-labels}, that the sensors have different data distributions. 
The low-cost sensor is detecting lower CO$_2$ \textcolor{black}{ppm} values which is unlikely to be true. This difference is likely due to the factory calibration of the sensor and the calibration drift these low-cost sensors experience. 
This is reinforced when viewing the change of the distribution of the readings for \textcolor{black}{Data Set} 2, seen in Figure \ref{fig:Hist-2-set2-features}, when compared to the change between the readings from Cape Point, seen in Figure \ref{fig:Hist-2-set2-labels}. It is interesting to note the shapes of the data distribution when comparing the data slices seen in Figures \ref{fig:Hist-4-set1-features} and \ref{fig:Hist-4-set2-features} when compared to their truth data seen in Figures \ref{fig:Hist-4-set1-labels} and \ref{fig:Hist-4-set2-labels}. The shapes of the distributions of the slices do not appear to change nearly as drastically in the truth data sets when compared to the change of the slices in the low-cost sensor data. This again further indicates sensor drift is present as well as the need for calibration.

These changes in distribution show that the sensor readings have indeed drifted to lower values over the months of testing while the values of the Cape Point sensor have not drifted appreciably. This drift indicates the need for more regular manual calibration of this low-cost sensor to trust the data collected. We aim to assist in extending the lifetime of manual calibration events through the use of machine learning which will be investigated further in future work.

Looking at the Shapiro-Wilk and Lilliefors' tests performed on the data slices in Tables \ref{tab:Lilliefors-2feat-2}-\ref{tab:Lilliefors-2label-4} we can see that the tests return P-Values that suggest that the null-hypothesis that the data is taken from a normal distribution. This means that we will need to consider this property of non-Gaussianity when doing future work on this problem. The statistic values from Shapiro-Wilk tests show that the truth data set consistently scores higher indicating that the data is closer to a normal distribution than that of the collected data set even though the P-value indicates that the null-hypothesis is rejected. When comparing the results of the Lilliefors' test, we see that the values of the Kolmogorov–Smirnov test statistic are smaller for the truth data set than that of the collected set. This further indicates that the collected data is likely less normally distributed than that of the truth data even though the P-value would indicate the test rejects the null-hypothesis that the data is from a normal set.

\section{Machine Learning Models for Calibration}
In this section the machine learning models will be investigated to perform calibration for our data. 
The models will be explained, briefly, and their hyper-parameters will be discussed. \textcolor{black}{Due to the low volume of data available, and the simplicity of the models, it was unfeasible to implement meaningful regularisation to these models.}
Each model is given \textcolor{black}{six} values within a 60 second window of the value it will predict as an input to allow for a one-minute prediction interval. The data is pre-processed to ensure each set of six values has a single truth data value for training as well as a time stamp. Any window of values around a single truth data point that had less than six values but more than three had the remaining empty values populated with the mean of the existing values to increase the size of the training data. Any window with three or less values is discarded and removed from the training and test data. Both \textcolor{black}{Data Set} 1 and 2 have been pre-processed in this fashion. Each model is trained with a 20\% validation split.

\subsection{Experimental Setup}

\subsubsection{Random Forest Regression}
Random Forests for regression \textcolor{black}{(RFR)} are formed by growing trees depending on a random vector that take on numerical values \cite{breiman2001random}. The result returned by the Random Forest is none other than the average of the numerical result returned by the different trees \cite{iannace2019wind}. \textcolor{black}{The model chosen used 10 estimators, selected due to the diminishing returns of estimators seen in \cite{conference2023}, and was trained with bootstrapping.}

\subsubsection{Support Vector Machine Regression}
Support Vector Regression (SVR) is a machine learning technique in which a model learns a variable's importance for characterizing the relationship between input and output \cite{zhang2020support}. One of the major tricks of Support Vector Machine (SVM) learning is the use of kernel functions to extend the class of decision functions to the non-linear case. This is done by mapping the data from the input space into a high-dimensional feature space as noted by R\"{u}ping\cite{Ruping2001kernels}. This made it an obvious candidate for testing and was implemented using a radial basis function (rbf) kernel as it performs well at a wide range of problems\cite{Ruping2001kernels}.

\subsubsection{1D Convolutional Neural Network}
Two dimensional Convolutional Neural Networks (CNN) have become one of the most representative neural networks in the field of deep learning \cite{li2021survey}. The one-dimensional Convolutional Neural Network (1D-CNN) is more suited to the one-dimensional data such as the data we have collected as seen in the work done by Rasheed et al \cite{rasheed2020improving} for predicting stock prices. It has also been used for spectroscopic signal regression in conjunction with other regression methods \cite{malek2018one}. This made it a candidate for our testing. The model was trained with the Adam optimizer and has six layers of 32 filters with scaling dilation rates of 1, 2, 4, 8, 16, 32 respectively and a single dense neural network output node.

\subsubsection{1D-CNN Long Short-Term Memory Network}
The 1D-CNN Long Short-Term network (LSTM) was shown to be more effective at predicting the remaining useful life of machining tools than that of the 1D-CNN or the LSTM individually \cite{niu2019remaining}. As we are already testing the 1D-CNN it was a natural choice to select the architecture for testing. This model was trained using the SGD optimizer. The model has one 1D-convolution with 32 filters followed by two LSTM layers with 32 nodes each and a single dense neural network output node. 

\subsection{Results}
\begin{figure}[!h]
    \centering
    \includegraphics[width=\columnwidth]{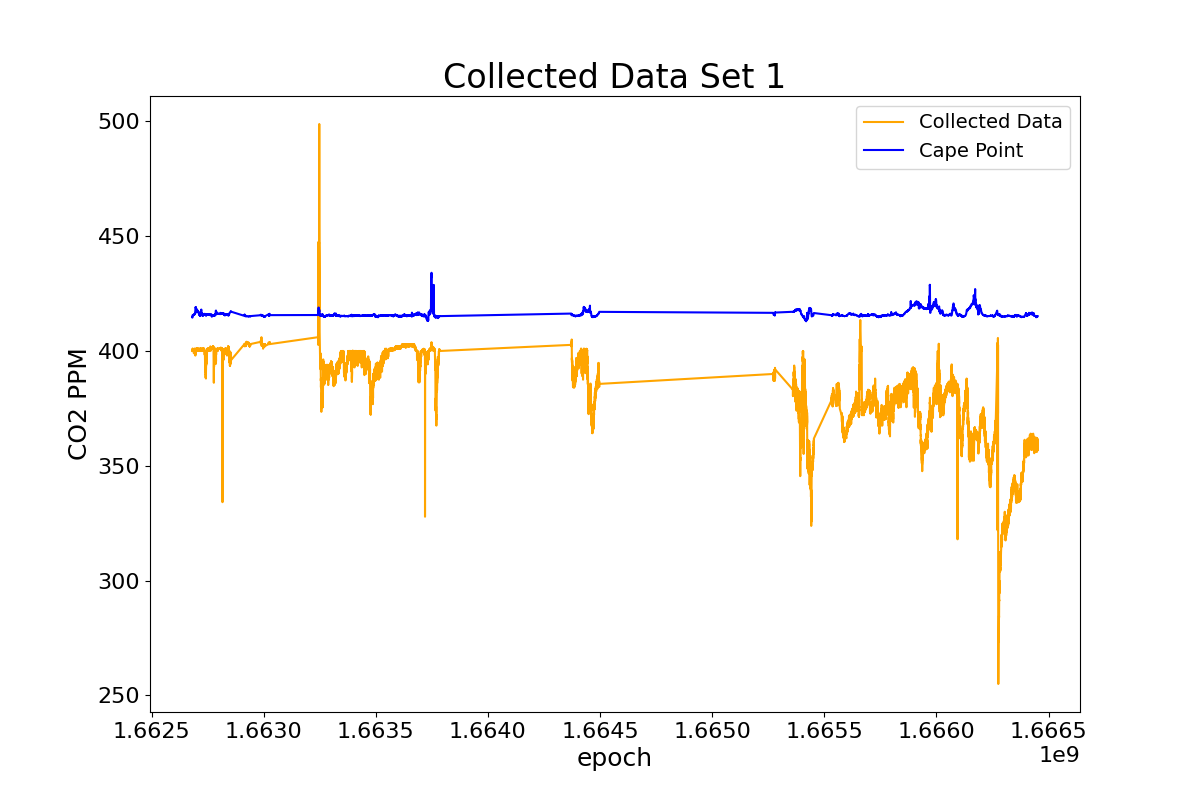}
    \caption{This plot shows the values of the collected \textcolor{black}{Data Set 1} compared to the Cape Point truth data. Note the epoch values are Unix epoch timestamps.}
    \label{fig:Collected-set1}
\end{figure}

\begin{figure}[!h]
    \centering
    \includegraphics[width=\columnwidth]{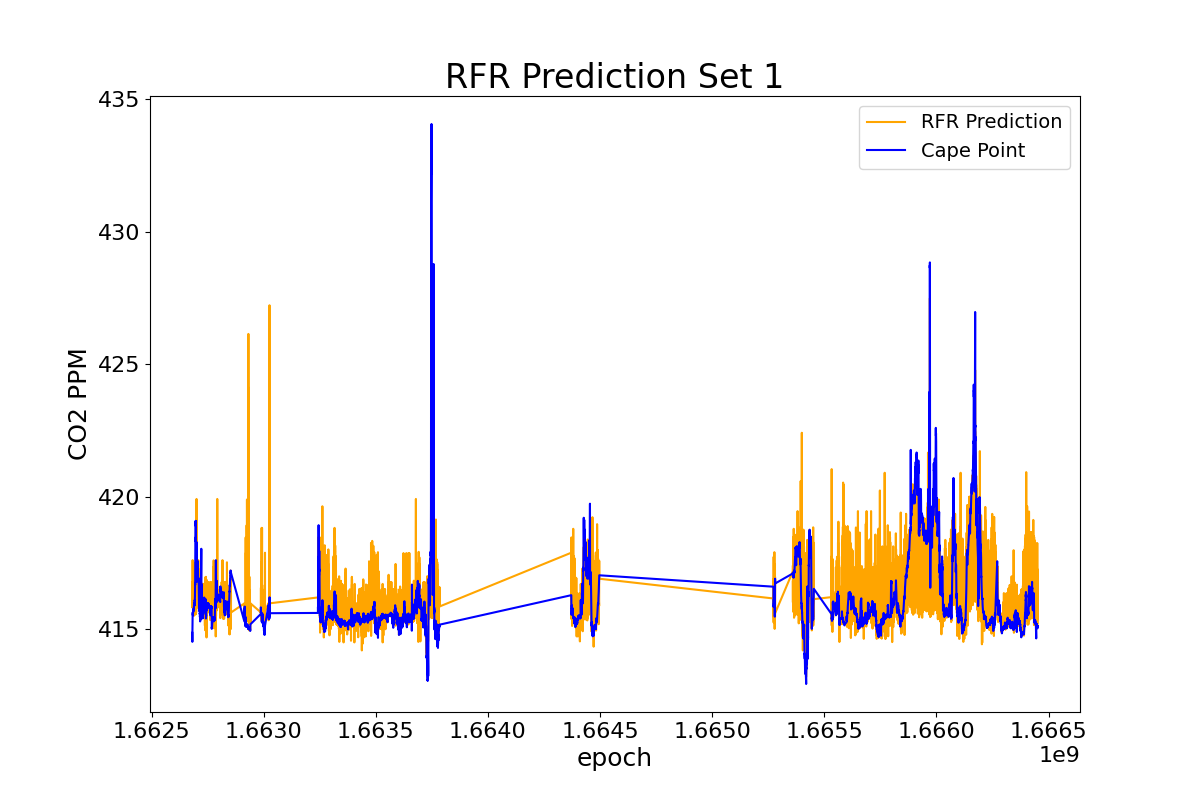}
    \caption{This plot shows the values predicted by the RFR model using the collected \textcolor{black}{Data Set 1} compared to the Cape Point truth data. Note the epoch values are Unix epoch timestamps.}
    \label{fig:RFR-set1-predicts}
\end{figure}

\begin{figure}[!h]
    \centering
    \includegraphics[width=\columnwidth]{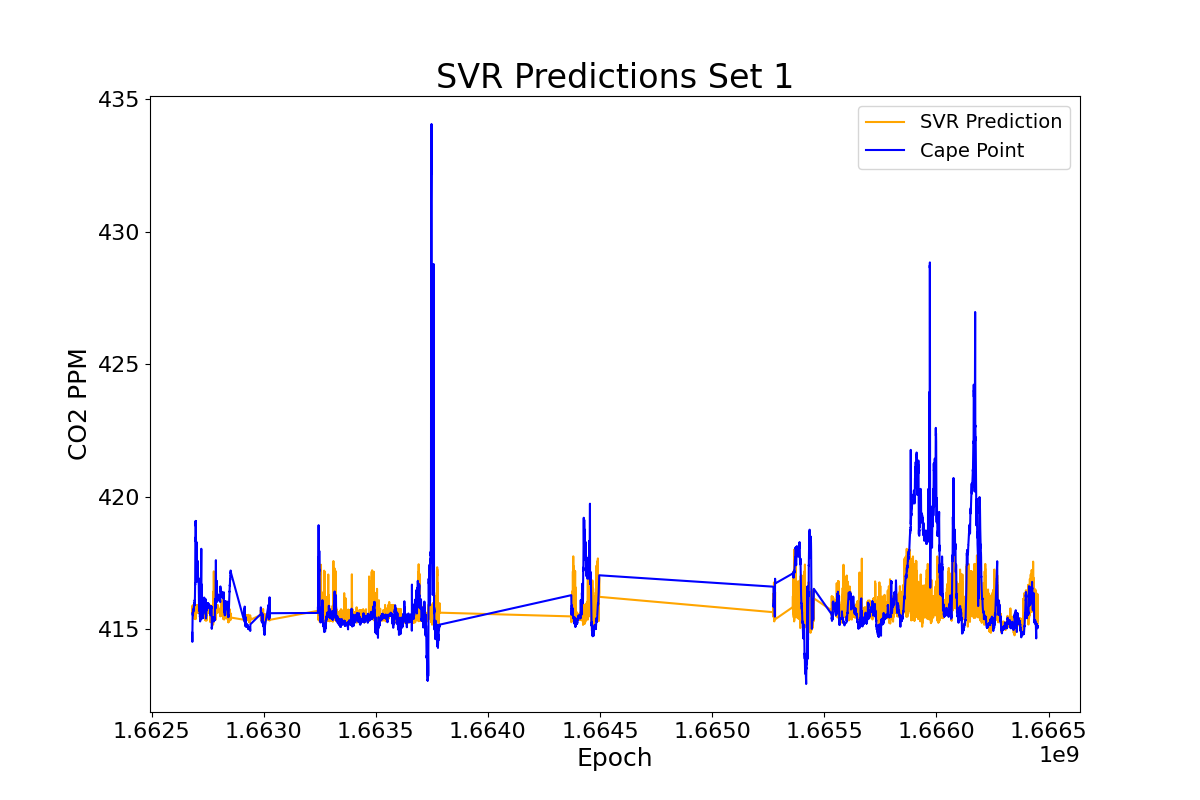}
    \caption{This plot shows the values predicted by the SVR model using the collected \textcolor{black}{Data Set 1} compared to the Cape Point truth data. Note the epoch values are Unix epoch timestamps.}
    \label{fig:SVR-set1-predicts}
\end{figure}

\begin{figure}[!h]
    \centering
    \includegraphics[width=\columnwidth]{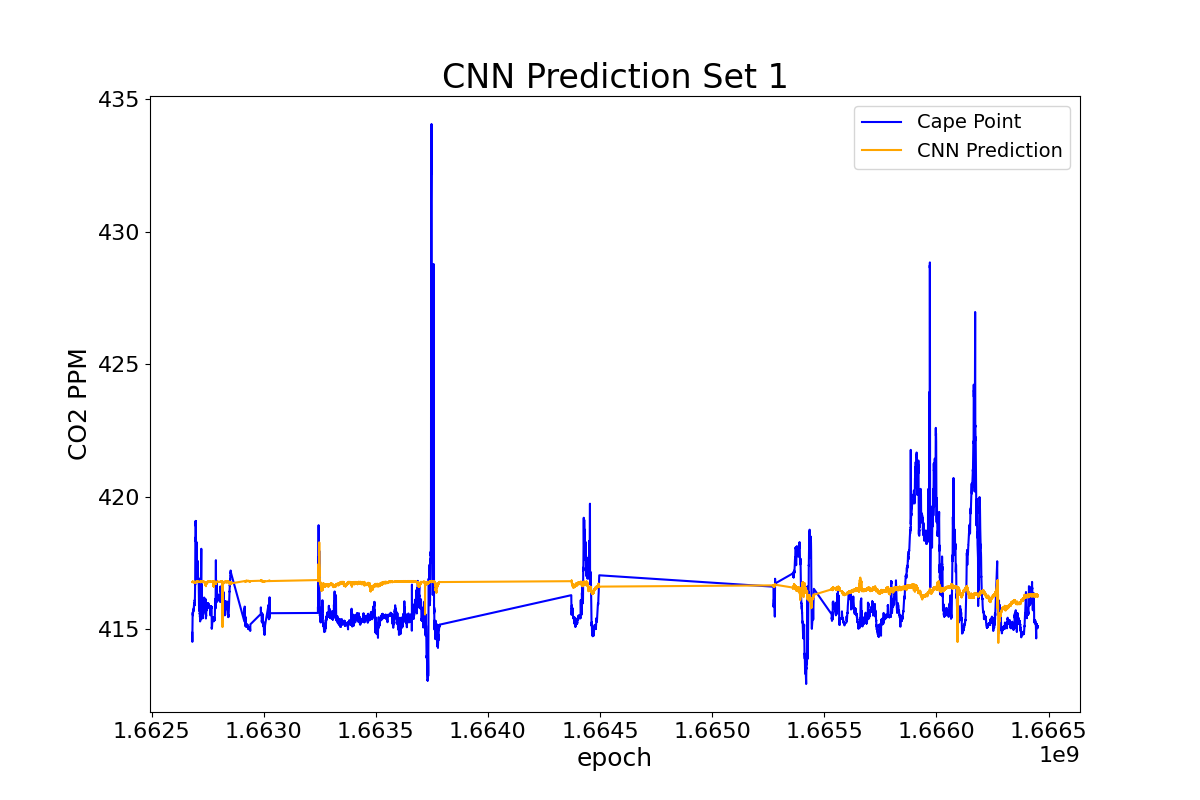}
    \caption{This plot shows the values predicted by the 1D-CNN using the collected \textcolor{black}{Data Set 1} compared to the Cape Point truth data. Note the epoch values are Unix epoch timestamps.}
    \label{fig:CNN-set1-predicts}
\end{figure}

\begin{figure}[!h]
    \centering
    \includegraphics[width=\columnwidth]{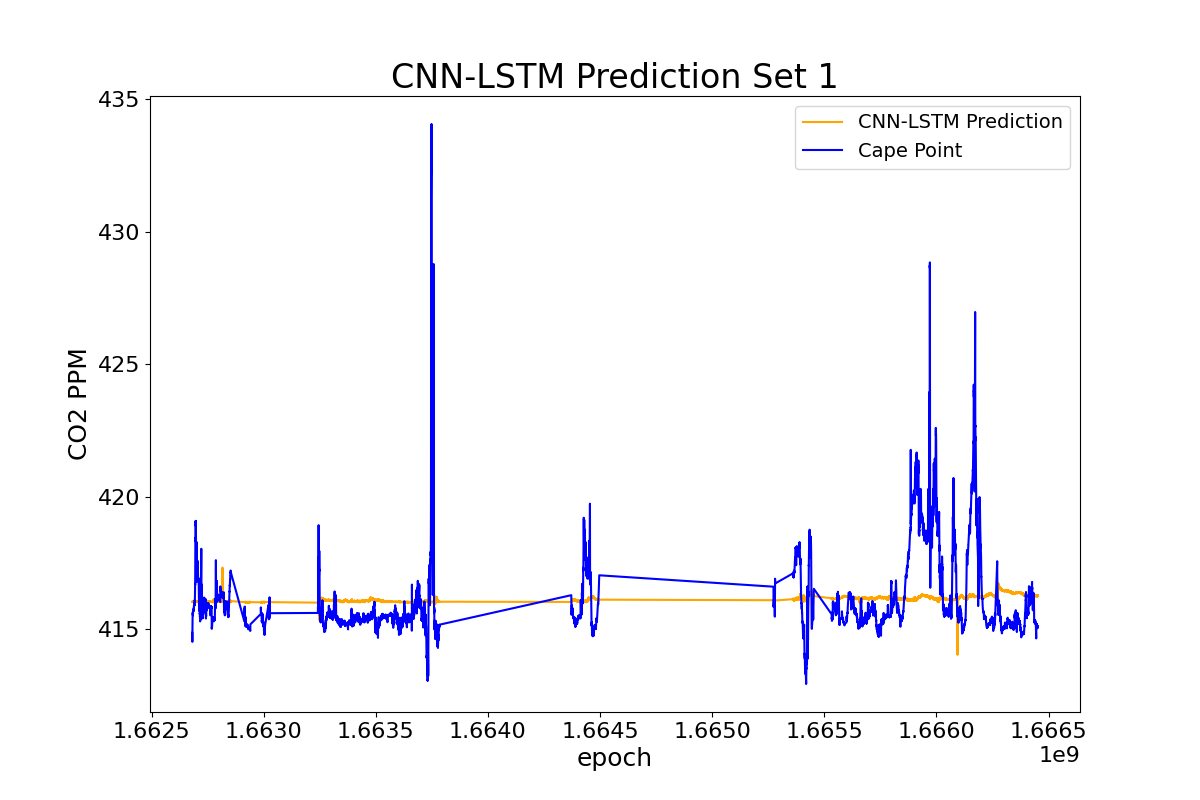}
    \caption{This plot shows the values predicted by the 1D-CNN LSTM using the collected \textcolor{black}{Data Set 1} compared to the Cape Point truth data. Note the epoch values are Unix epoch timestamps.}
    \label{fig:CNN-LSTM-set1-predicts}
\end{figure}

In this section we see the results of the prediction models when trained on set one. The raw collected data can be seen in Figure \ref{fig:Collected-set1} and can be compared to the predictions, using this data, made by the machine learning models. The predictions of the models can be seen in Figures \ref{fig:RFR-set1-predicts}-\ref{fig:CNN-LSTM-set1-predicts}.

\subsection{Discussion}
When looking at the collected data from the low-cost sensor, seen in Figure \ref{fig:Collected-set1}, it is clear that the sensor is not accurate. It is also evident that the sensor readings are drifting over time. This level of performance from the sensor would not be usable for any kind of data analysis into the changes of CO$_2$ concentration as the sensor is giving inaccurate values and is trending in a way that is not consistent with the truth data.
Looking at the predictions made by the different models using this collected data set, we can see the accuracy of the data has been increased. The RFR model predictions, seen in Figure \ref{fig:RFR-set1-predicts}, have the most variance of the models. This is then followed by the SVR model, seen in Figure \ref{fig:SVR-set1-predicts}, with the most similar level of variance to the truth data. Finally we see that both the 1D-CNN and the 1D-CNN LSTM models, seen in Figures \ref{fig:CNN-set1-predicts} and \ref{fig:CNN-LSTM-set1-predicts} respectively, have the lowest levels of variance in their predictions. This is likely due to the feature extraction properties of the CNN architecture. When looking at the truth data, the level of variation in the signal is significant and cannot be considered as random noise. This is due to the high level of sensitivity of the instrument when compared to the scale of the graphs presented. While, in many situations, the levels of variance of the RFR and SVR may be considered as over-fitting when compared to the 1D-CNN and the 1D-CNN LSTM, in this situation the variance of the truth data carries significant information due to the accuracy of the truth data instrument. Therefore, the variance of the SVR model being similar to that of the truth data is beneficial.

In future we would like to explore more models and combinations of architectures to improve the performance of these models. The results from these experiments have produced promising results, however due to the stability of the measurement site we feel as more work should be done on data with higher variability to show the performance of these models in different environments.

\section{Statistical and Ergodicity Study of the Performance of the Machine Learning Models}
In order to determine the performance of the machine learning models it was necessary to compare them across different metrics. The statistical performance of these models was investigated and compared. 

\subsection{Experimental Setup}

The models were compared with a variety of metrics that look both at the prediction error as well as the fit and the statistical similarity to the truth data. This will allow us to determine the performance of the model as its performance changes over time. To do this, some metrics of comparison have been chosen. These metrics are seen in Table \ref{tab:metrics}.

\begin{equation}
    MAE = (\frac{1}{n})\sum_{i=1}^{n}\left | \hat{y}_{i} - y_{i} \right |
\end{equation}

\begin{equation}
    Accuracy(\%) = 100 - \frac{1}{n}\sum_{i=1}^{n}(100\frac{\left | \hat{y}_{i} - y_{i} \right |}{y_{i}})
\end{equation}

\begin{equation}
    R^2 = 1 - \frac{(y_{i}-\hat{y}_{i})^{2}}{(y_{i}-\Bar{y})^{2}}
\end{equation}

\begin{equation}
     \rho = \frac{\text{cov}(\hat{Y},Y)}{\sigma_{\hat{y}} \sigma_y}
\end{equation}

\begin{equation}
    KL(\hat{y} || y) = \sum_{c=1}^{M}\hat{y}_c \log{\frac{\hat{y}_c}{y_c}}
\end{equation}

\begin{equation}
    JS(\hat{y} || y) = \frac{1}{2}(KL(y||\frac{y+\hat{y}}{2}) + KL(\hat{y}||\frac{y+\hat{y}}{2}))   
\end{equation}
 Where:\\ $y$ is the truth value\\$\hat{y}$ is the predicted value\\$\Bar{y}$ is the mean value\\



We have chosen to look at a range of features. This includes the divergence of the predicted values from the truth values in the form of Kullback-Liebler and Jenson-Shannon divergence. These values should give an idea of the difference of the distributions created by the model prediction and the truth data through the relative entropy of the predictions and the truth data. This should give insight into the statistical similarity of the predictions which may be relevant in certain fields of research on this type of data. Each model has been tested on both collected data sets for the predictions allowing for the performance of the model to be evaluated on its performance on data which has been seen in training and close to the training data in time as well as data that is not immediately within the same time frame as to determine the change in performance. The data in \textcolor{black}{Data Set 1} contains 30000 groups of six values while \textcolor{black}{Data Set 2} contains 8000 groups of six values. All models have then been trained on each set and have predicted the truth data of the same set as well as the truth data of the other set.
 
\subsection{Results}
The results of these experiments can be seen in Table \ref{tab:ML-Model-Stats}.
\begin{table*}[]
\caption{Table showing the scores of different the machine learning models tested indicating their performance at predicting values and their statistical similarity to the truth data.}
\resizebox{\textwidth}{!}{%
\begin{tabular}{|l|l|l|l|l|l|l|l|}
\hline
\textbf{Train:Predict} & \textbf{Model} & \textbf{MAE (\textcolor{black}{ppm})} & \textbf{Accuracy (\%)} & \textbf{R$^2$}          & \textbf{Pearson}      & \textbf{Kullback-Liebler} & \textbf{Jenson-Shannon} \\ \hline
\textcolor{black}{Data Set 1:Data Set 1}                    & RFR            & 0.89               & 99.79                  & 0.243  & 0.493   & 0.266        & 0.058    \\
                       & CNN            & 1.33               & 99.68                  & -0.090 & -0.018 & 0.413        & 0.101     \\
                       & CNN-LSTM       & 1.34               & 99.68                  & -0.341               & -0.043                & 0.379                     & 0.092                   \\
                       & \textcolor{black}{\textbf{SVR}}            & \textcolor{black}{\textbf{0.87}}               & \textcolor{black}{\textbf{99.79}}                  & \textcolor{black}{\textbf{0.012}} & \textcolor{black}{\textbf{0.315}}   & \textcolor{black}{\textbf{0.151}}       & \textcolor{black}{\textbf{0.037}}   \\ \hline
\textcolor{black}{Data Set 2:Data Set 2}                    & RFR            & 0.45               & 99.89                  & 0.271  & 0.520    & 0.376       & 0.095     \\
                       & CNN            & 0.92               & 99.78                  & -1.030  & -0.074    & 0.528        & 0.135     \\
                       & CNN-LSTM       & 0.67               & 99.84                  & -0.450               & -0.060                & 0.250                     & 0.061                   \\
                       & \textcolor{black}{\textbf{SVR}}            & \textcolor{black}{\textbf{0.49}}               & \textcolor{black}{\textbf{99.88}}                  & \textcolor{black}{\textbf{0.077}}   & \textcolor{black}{\textbf{0.279}}    & \textcolor{black}{\textbf{0.341}}       & \textcolor{black}{\textbf{0.082}}     \\ \hline
\textcolor{black}{Data Set 1:Data Set 2}                    & RFR            & 0.63               & 99.85                  & -0.114 & 0.067   & 0.290       & 0.076     \\
                       & CNN            & 0.65               & 99.84                  & -0.319  & -0.069   & 0.485        & 0.122    \\
                       & CNN-LSTM       & 2.64               & 99.37                  & -11.706              & -0.068                & 0.274                     & 0.065                   \\
                       & \textcolor{black}{\textbf{SVR}}            & \textcolor{black}{\textbf{0.62}}               & \textcolor{black}{\textbf{99.85}}                  & \textcolor{black}{\textbf{-0.095}} & \textcolor{black}{\textbf{0.076}}   & \textcolor{black}{\textbf{0.152}}       & \textcolor{black}{\textbf{0.036}}    \\ \hline
\textcolor{black}{Data Set 2:Data Set 1}                    & RFR            & 1.05               & 99.75                  & -0.284 & -0.090  & 0.536        & 0.129       \\
                       & CNN            & 1.24               & 99.70                  & -0.170 & -0.073  & 0.296       & 0.071     \\
                       & CNN-LSTM       & 1.89               & 99.55                  & -0.621               & -0.042                & 0.394                     & 0.100                   \\
                       & \textcolor{black}{\textbf{SVR}}            & \textcolor{black}{\textbf{0.96}}               & \textcolor{black}{\textbf{99.77}}                  & \textcolor{black}{\textbf{-0.125}}  & \textcolor{black}{\textbf{-0.047}}  & \textcolor{black}{\textbf{0.518}}        & \textcolor{black}{\textbf{0.132}}     \\ \hline
\end{tabular}%
}
\label{tab:ML-Model-Stats}
\end{table*}

\subsection{Discussion}
As seen in Table \ref{tab:ML-Model-Stats}, the models have been tested and show an improvement over the collected values. We are aware that the final test group where the models have been trained on future data and are predicting past values is non-causal however, it has been included as the models do not use the time labels of the data. Looking at the mean absolute error we find that the RFR and the SVR are consistently producing lower values than the 1D-CNN and CNN-LSTM models. This also results in consistently higher accuracy scores for both the RFR and SVR models on these data sets. It is interesting to note that the mean absolute error of the predictions are lower when predicting \textcolor{black}{Data Set 2 when trained on Data Sets 1 and 2 than when predicting Data Set 1. This is likely due to the length of Data Set 2 when compared to Data Set 1. The only exception to this is the CNN-LSTM when predicting Data Set 2 when trained on Data Set 1.} When looking at the R$^2$ scores we see a generally weak correlation across the models. This is further confirmed by the Pearson Correlation Coefficient. The highest scores for the Pearson Correlation Coefficient are seen in the RFR models performance on the same set as it was trained on. Comparing the Kullback-Liebler and Jenson-Shannon divergence scores, we see that the SVR model performs the best on all the causal tests. This is closely followed by the RFR model which is consistently second in the causal tests. We do see the best performance in test two of the CNN-LSTM model. Interestingly the CNN based models performed better in this metric in the final non-causal test than the RFR and SVR models. For these data sets the overall best performing model is the SVR as it scores very closely with the RFR but performs better in the Kullback-Liebler and Jenson-Shannon divergence tests indicating the closest recreation of the truth data. From the results we see that the prediction performance has not appreciably degraded when predicting \textcolor{black}{Data Set 2, which starts on the 27/01/2023 and ends on 02/02/2023, when trained on Data Set} 1, which spans 07/09/2022 to 22/10/2022, indicating the models would still be suitable for use after two months of separation between training and prediction at this site. 
In future we would like to test over a longer interval between training and prediction. Our co-located site is still operational, and this data will be available for future testing. We plan to test on a more variable data set as the set we have collected for testing is very stable. We are working on expanding our data collection to different locations and sensor technologies to evaluate the models in different scenarios and operating conditions. We would also like to make use of different techniques with or separately from the work presented to improve the performance of these, and other, automatic calibration models.

\section{\textcolor{black}{Long-term Testing}}
\color{black}
In this section we inspect the performance of the models trained on \textcolor{black}{Data Set} 1 and 2 when applied to a longer period of unseen data. This will give an idea of the long-term robustness of the models.

\subsection{Experimental Setup}
\begin{figure}[!h]
    \centering
    \includegraphics[width=\columnwidth]{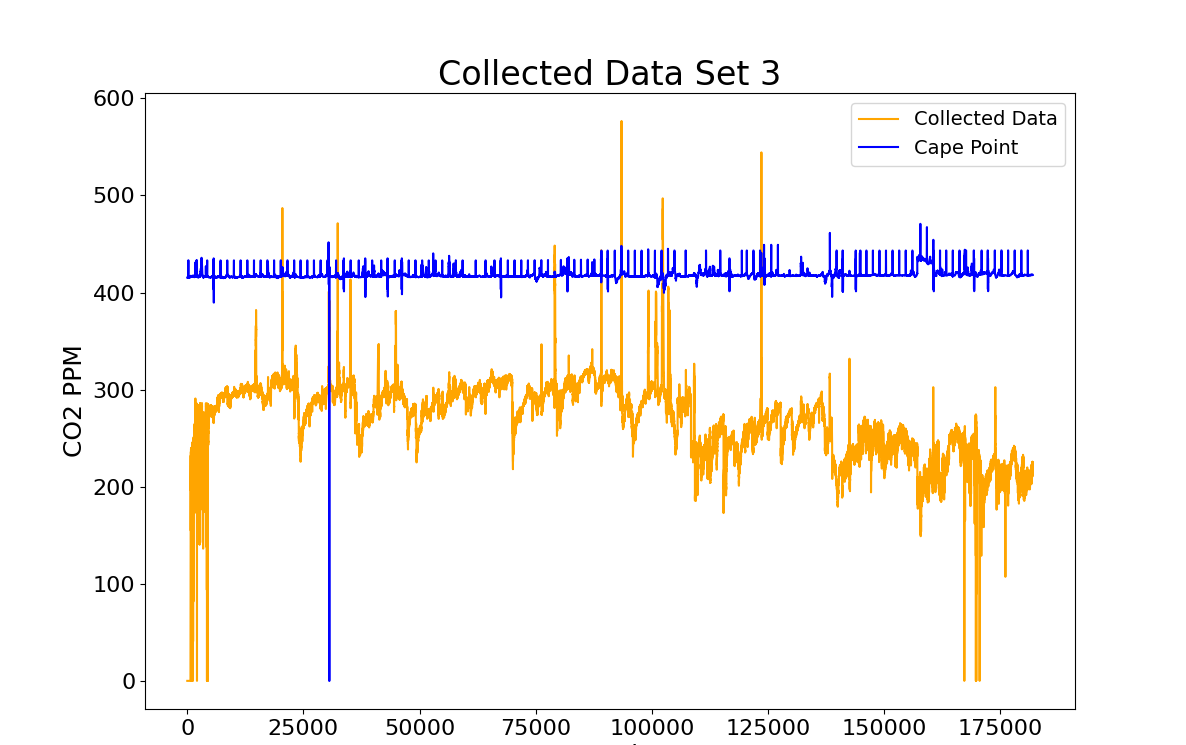}
    \caption{This plot shows the collected CO$_2$ \textcolor{black}{ppm} and truth data CO$_2$ \textcolor{black}{ppm for Data Set} 3 from 18/03/2023 to 25/10/2023.}
    \label{fig:Set3Raw}
\end{figure}

A large data set, \textcolor{black}{Data Set} 3, collected from the Cape Point measurement site spanning 18/03/2023 to 25/10/2023 will be used to asses the models. This data set is comprised of 182000 truth data points and 1092000 windowed, matched, collected data points. To verify the long-term performance of the models, each models trained on \textcolor{black}{Data Set} 1 and 2 will be predict the data in this new \textcolor{black}{Data Set} 3. This data set can be seen in Figure \ref{fig:Set3Raw}. This data set appears to show the long term drift of the low-cost sensor.

\subsection{Results}
The results of this experiment can be seen in Table \ref{tab:ML-Model-Stats-set3}.
\begin{table*}[]
\caption{Table showing the scores of different the machine learning models tested indicating their performance at predicting values and their statistical similarity to the truth data.}
\resizebox{\textwidth}{!}{%
\begin{tabular}{|l|l|l|l|l|l|l|l|}
\hline
\textbf{Train:Predict} & \textbf{Model} & \textbf{MAE (\textcolor{black}{ppm})} & \textbf{Accuracy (\%)} & \textbf{R$^2$}          & \textbf{Pearson}      & \textbf{Kullback-Liebler} & \textbf{Jenson-Shannon} \\ \hline
\textcolor{black}{Data Set 1:Data Set 3}                    & RFR            & 2.79               & 36.76                  & -0.101  & 0.026   & 0.585        & 0.145    \\
                       & CNN            & 3.79               & 36.92                  & -0.262 & -0.109 & 0.513        & 0.121     \\
                       & CNN-LSTM       & 3.04               & 36.53                  & -0.016               & 0.131                & 0.516                     & 0.122                   \\
                       & \textcolor{black}{\textbf{SVR}}            & \textcolor{black}{\textbf{2.77}}               & \textcolor{black}{\textbf{36.79}}                  & \textcolor{black}{\textbf{-0.098}} & \textcolor{black}{\textbf{0.047}}   & \textcolor{black}{\textbf{0.545}}       & \textcolor{black}{\textbf{0.130}}    \\ \hline
\textcolor{black}{Data Set 2:Data Set 3}                    & RFR            & 2.98               & 36.82                  & -0.119  & 0.053    & 0.525       & 0.132     \\
                       & CNN            & 3.93               & 37.12                  & -0.058  & 0.130    & 0.577        & 0.139     \\
                       & CNN-LSTM       & 2.99               & 37.16                  & -0.058               & 0.130                & 0.518                     & 0.122                   \\
                       & \textcolor{black}{\textbf{SVR}}            &   \textcolor{black}{\textbf{3.00}}             & \textcolor{black}{\textbf{36.82}}                  & \textcolor{black}{\textbf{-0.123}}   & \textcolor{black}{\textbf{0.038}}    & \textcolor{black}{\textbf{0.491}}       & \textcolor{black}{\textbf{0.115}}     \\ \hline

\end{tabular}%
}
\label{tab:ML-Model-Stats-set3}
\end{table*}

\begin{table}[!h]
\caption{Train-Predict matrix of MAE (\textcolor{black}{ppm}) for Random Forest Regression (RFR) model}
\label{tab:Matrix-RFR}
\resizebox{\columnwidth}{!}{%
\begin{tabular}{l|lll|}
\cline{2-4}
                                        & Predict: Data Set 1 &Predict: Data Set 2 &Predict: Data Set 3 \\ \hline
\multicolumn{1}{|l|}{Train: Data Set 1} & 0.89       & 0.63       & 2.79       \\
\multicolumn{1}{|l|}{Train: Data Set 2} & 1.05       & 0.45       & 2.98       \\
\multicolumn{1}{|l|}{Train: Data Set 3}  & 12.89      & 3.07       & 1.98       \\ \hline
\end{tabular}%
}
\end{table}

\begin{table}[]
\caption{Train-Predict matrix of MAE (\textcolor{black}{ppm}) for Support Vector Regression (SVR) model}
\label{tab:Matrix-SVR}
\resizebox{\columnwidth}{!}{%
\begin{tabular}{l|lll|}
\cline{2-4}
                                        &Predict: Data Set 1 &Predict: Data Set 2 &Predict: Data Set 3 \\ \hline
\multicolumn{1}{|l|}{Train: Data Set 1} & 0.87       & 0.62       & 2.77       \\
\multicolumn{1}{|l|}{Train: Data Set 2} & 0.96       & 0.49       & 3.00       \\
\multicolumn{1}{|l|}{Train: Data Set 3} & 2.69       & 1.64       & 1.95       \\ \hline
\end{tabular}%
}
\end{table}

\subsection{Discussion}
From the results seen in Table \ref{tab:ML-Model-Stats-set3} we can see that the performance of the models has degraded when compare to the performance seen in Table \ref{tab:ML-Model-Stats}. This is to be expected due to the relatively large difference in time between the model calibration and the data used in this experiment. It is clear that, while significantly better than uncalibrated, the models do not achieve the same levels of performance over this period of time. We can see that the MAE of the models have increased  while percentage accuracy has greatly decreased. The R$^2$ and Pearson correlation coefficient scores indicate little correlation between the predictions of the models and the truth data. This experiment also shows that over time the statistical similarity of the predictions degrade showing significantly worse scores for all the models in the Kullback-Liebler and Jenson-Shannon divergence. It is clear from this experiment that additional factors may need to be considered to improve the robustness of these models. In Tables \ref{tab:Matrix-RFR} and \ref{tab:Matrix-SVR} we see the MAE scores for training and prediction across the different data sets. It shows that the training on a shorter period of data produces better results. Further investigation into the relationship between the prediction error and the training period will be addressed in future work.

\color{black}
\section{Conclusion}
Inexpensive and low-fidelity sensors would be a major enabler in making sensing ubiquitous which would ultimately empower us in various ways to achieve the sustainable development goals. 
Data quality is a major issue for such inexpensive sensors. In our work, we have presented a thorough analysis of statistical characterization of data from one such inexpensive sensor and have investigated a set of calibration algorithms. 

We found that the data collected from our low-cost sensor produced a non-normal distribution while the truth data from the site appeared to have a more normal distribution due to the statistic values from the Shapiro-Wilk and Lilliefors' tests, however this could not be confirmed as our P-values from both the Shapiro-Wilk and Lilliefors' tests rejected the null-hypothesis that the data was drawn from a normally distributed set. This indicates that we may need to investigate options for automatic calibration that takes this property of non-Gaussianity into account. We also noted the calibration drift of the low-cost sensor as the data collected was trending in a manner that was not consistent with the truth data from the site. We trained RFR, SVR, 1D-CNN and 1D-CNN LSTM models on our data sets by using six time-widowed values to predict a one-minute interval CO$_2$ concentration with matched truth data. We found that the predictions improved the performance of the sensor beyond its raw readings. 
It was noted that the SVR model had a more similar level of variation to the truth data when compared to the other models tested. Our work has found that for this data set the SVR model has been shown to be the overall best choice for automatic calibration from our results. \textcolor{black}{This is a different result to the work done in our previous paper \cite{conference2023}, however due to the larger data sets used in this analysis it was likely that a model with more robustness would perform better over a larger period of time. It was found that the performance of all the models degrade over time. This indicates that there are factors that these machine learning models are not able to account for over a larger period of time. The models did, however, in this case improve the mean absolute error of the low-cost sensor significantly over the raw captured values.} The results from our testing suggests that it may be possible to use these models to improve the accuracy and manual calibration lifetime of low-cost sensors, such as the one used in this study, to allow for more cost-effective wide-scale air quality monitoring systems to be effectively implemented.

A limitation of this work is due to the stability of our testing site as this may impact the performance of the machine learning models as this may artificially increase their performance metrics. 
This is a limitation in many parts of the Global South. In future, we plan to study this in detail and work on solutions which can bolster the systems. 
Another limitation is that we are not considering other factors that may influence the performance of the sensor such as pressure, humidity or temperature. \textcolor{black}{In future work we may include these and other factors into new models by making use of physics based models and machine learning models with the aim of increasing their performance and robustness. Further investigation into the relationship between prediction error and training period is needed to improve the performance of the machine learning models.} With this work we aim to work towards implementing our patent that has been registered with the UK patent office with the title, “Method and System of Calibration of a Sensor or a network of Sensors”, and application number 2215800.0 (filed on 25-Oct-2022).
 In addition to environmental sensors, we also plan to work on calibration challenges faced by industrial sensors as well. 

\section*{Acknowledgment}
We would like to thank Sentech Soc Ltd, Weather South Africa and their team at Cape Point for their assistance with our research. We acknowledge Sentech Plc whose funding for the Centre for 5G for Sustainable Development has funded this work. We also acknowledge the anonymous reviewers whose constructive feedback has bolstered this paper.

\bibliographystyle{IEEEtran}
\bibliography{Bib}


\begin{IEEEbiography}[{\includegraphics[width=1in,height=1.25in,clip,keepaspectratio]{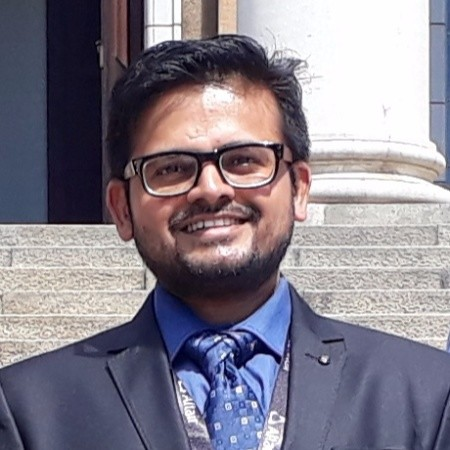}}]{AMIT KUMAR MISHRA (Senior Member, IEEE)} is an active researcher in the domain of sensor design, radar, applied machine learning, and frugal innovation. He is currently a Professor with the Department of Electrical Engineering, University of Cape Town (UCT), Cape Town, South Africa. He heads the Sentech funded Center of Excellence in 5G for Sustainable Development at UCT. His Google Scholar-based H-index is 19. He has more than 150 peer-reviewed publications and holds five patents. He also holds a joint appointment as the Professor of Sensors and AI at University West in Sweden.
\end{IEEEbiography}

\begin{IEEEbiography}[{\includegraphics[width=1in,height=1.25in,clip,keepaspectratio]{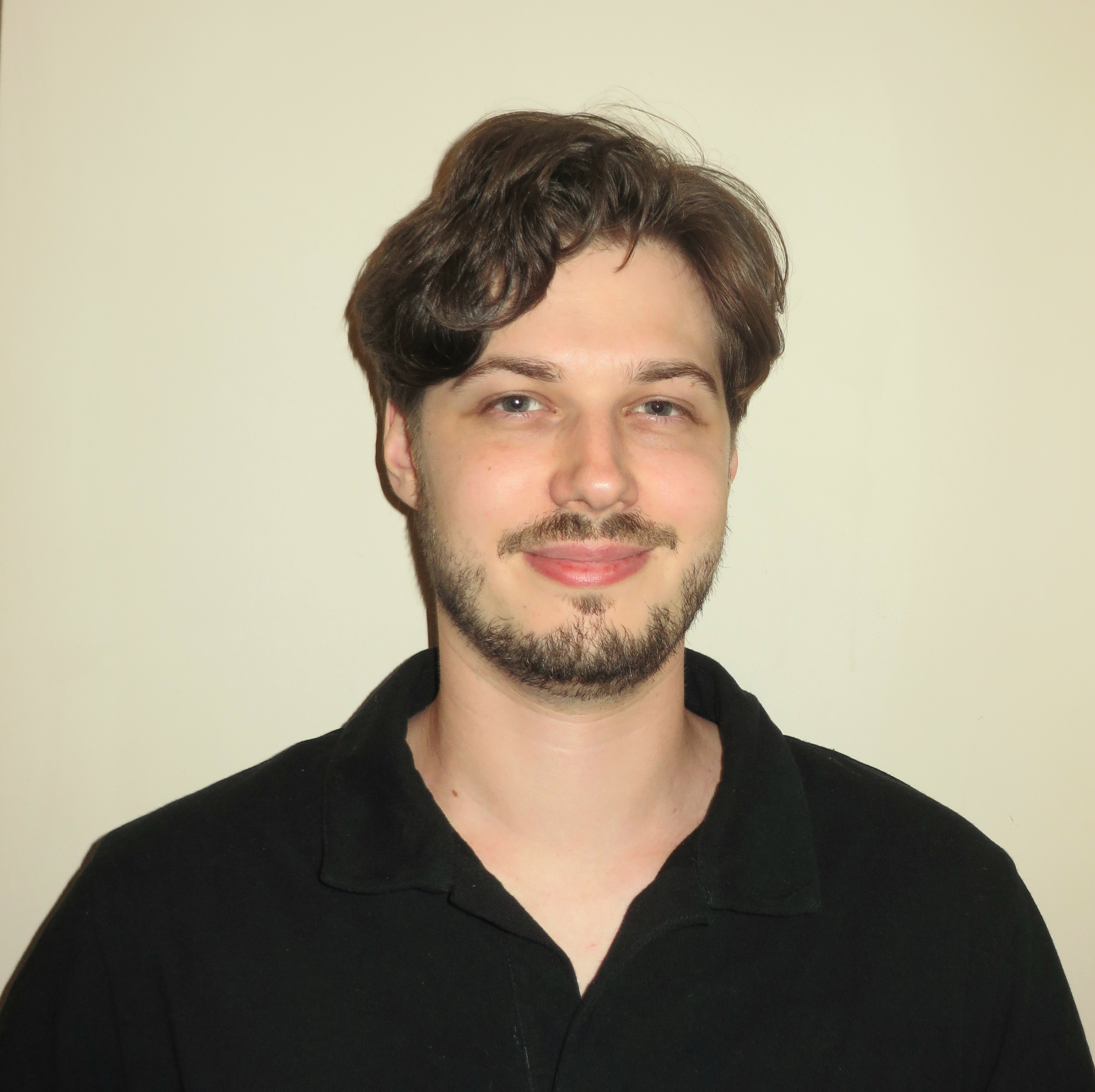}}]{TRAVIS BARRETT} 
is currently a postgraduate research scholar in the Department of Electrical Engineering, University of Cape Town (UCT), Cape Town, South Africa. His research is in the domain of cost-effective environment sensing embedded systems and automatic sensor calibration.
\end{IEEEbiography}

\end{document}